\documentclass[12pt,a4paper]{article}
\usepackage[latin1]{inputenc}
\usepackage{amsmath}
\usepackage{amsfonts}
\usepackage{amssymb}
\usepackage{graphicx}
\usepackage{subfigure}
\usepackage{cite}
\newcommand{\angstrom}{\textup{\AA}}

\begin{document}

\date{}

\author{ Sunil K. Tripathy\footnote{tripathy\_ sunil@rediffmail.com}  and  Anup Pattanaik\footnote{anup.pattanaik@igitsarang.ac.in} \\
Department of Physics,\\
Indira Gandhi Institute of Technology, Sarang\\
Dhenkanal, Odisha-759146, INDIA}

\title  {\bf  Optical and electronic properties of some binary semiconductors from energy gaps}

\maketitle

\begin{abstract}
II-VI and III-V tetrahedral semiconductors have significant potential for novel optoelectronic applications. In the present work, some of the optical and electronic properties of these groups of semiconductors have been studied using a recently proposed empirical relationship for refractive index from energy gap. The calculated values of these properties are also compared with those calculated from some well known relationships. From an analysis of the calculated electronic polarisability of these tetrahedral binary semiconductors from different formulations, we have proposed an empirical relation for its calculation. The predicted values of electronic polarisability of these semiconductors agree fairly well with the known values over a wide range of energy gap.
\end{abstract}


\section{Introduction}

Semiconductors rich in optoelectronic properties have generated a lot of research and technological interest in recent times because of their novel applications in  different devices such as light emitting diodes (LED), laser diodes (LD), integrated circuits (IC), photo detectors (PD), nanotechnology, heterostructure lasers and optical modulators operating in mid infra-red regions ($2-5 \mu m$) \cite{Paskov97,Rappl01}. Among different novel materials fabricated, II-VI and III-V group semiconductors and alloys find wide range of interesting applications. These materials exhibit Wurzite and Zinc blende tetrahedral crystal structures. Wide band gap II-VI semiconductors are used in optoelectronic devices such as LEDs and LDs operating in blue-green spectral range \cite{Kasap07}. These materials are characterized by different degree of ionicity which make them suitable for high electro-optical and electromechanical coupling.  II-VI binary semiconductors such as ZnS, ZnSe and ZnTe find applications as blue lasing materials and can be used in the fabrication of optical wave guides and modulated heterostructure \cite{Khenata06,Hasse91, Tamargo91}. Oxides like ZnO are used in nano medicines \cite{Grone06}. Recently there have been a lot of studies on the elastic, electronic and optical properties of some II-VI group semiconductors \cite{Khenata06, Reshak06, Reshak07, AlDouri10, Abdul12, Trojnar12, Lippens91, Lefeb94, Umar12, Reshak14,Wong13, Wong13a, Reynolds65}. III-V group of semiconductors also exhibit tetrahedral structures and have interesting applications in optoelectronics and photovoltaics because of their direct band gaps and high refractive indices. They are used in the fabrications of high efficiency solar cells. Large breakdown fields, high thermal conductivities and electron transport properties of III-V nitrides such as GaN, InN, AlN make them suitable for novel optoelectronic applications in visible and ultra violet spectral range \cite{Kasap07}. In recent times there have been a lot of interest in the calculation of the electronic and optical properties of III-V binary semiconductors and their alloys \cite{Johnson90,Reshak06a,AlDouri11, Bredin94, Borak05, Levine91, Reshak07a, Reshak05,Tit10,Haq14, Aly15}.

In semiconductors, two basic properties namely energy gap and refractive index mostly decide their optical and electronic behaviour. Refractive index of a material usually decreases with an increase in energy gap fostering an underlying relationship between these two fundamental quantities. There have been many attempts to correlate these two quantities with a suitable empirical or semi-empirical relation [31-48]. Recently, an empirical relation, based on some experimental refractive index data of some elemental and binary semiconductors has been proposed by Tripathy, which can be equally applied to all regions of energy gap and can also be applied to ternary semiconductors \cite{SKT15}. In the present work, we have used this recently proposed relation to investigate some of the optical and electronic properties of II-VI and III-V group semiconductors.  The organisation of the work is as follows. In Sect-2, we  have calculated different optical properties of II-VI and III-V group semiconductors such as dielectric constant, optical susceptibility and reflectivity using the relation proposed in Ref.\cite{SKT15}. The calculated values are also compared with the values predicted using some well known relations. In Sect-3, different electronic properties such as ionicity and electronic polarisability of III-V and II-VI group semiconductors have been calculated and compared with the predicted values of other calculations. From an analysis of the electronic polarisability of the semiconductors, we have proposed an empirical relation for its calculation. At the end, conclusions and summary of the work are presented in Sect-4.

\section{Optical Properties }
\subsection{\it Refractive Index}

Refractive index $n$ of a material is a measure of its transparency to the incident photons. For semiconductor materials, it is considered as an important physical parameter. A proper design of optoelectronic device needs an accurate knowledge of refractive indices of materials. Refractive index is closely related to the electronic properties and band structure of the material. Theoretically, two different approaches are adopted to calculate the refractive index. In the first approach, it is calculated from the electronic behaviour concerning the local fields and the molar volume of the material. In the second approach, a more involved quantum mechanical calculation is performed to calculate the band structure which in turn is related to the refractive index of the material. Since the band structure of a semiconductor is intimately related to its optical behaviour, there have been attempts to calculate refractive index from energy gaps. In this context, so many empirical relations between refractive index and energy gap $E_g$ have been proposed earlier\cite{Moss50, Ravi79, Penn62,Reddy08, Reddy09, Herve94, Herve95, Kumar10}. These relations have been widely used in literature to calculate different opto-electronic properties of different groups of semiconductors. Recently,   Tripathy has proposed an empirical relation to calculate refractive index of semiconductors  from their energy gaps\cite{SKT15}. In that work, it has been shown that, the relation (hereafter termed as Tripathy relation) can be successfully used for different group of semiconductors for a wide range of energy gaps. According to the Tripathy relation, the refractive index of a semiconductor with energy gap $E_g$ is given by

\begin{equation}
n=n_0[1+\alpha e^{-\beta E_g}],\label{e1}
\end{equation}
where, $n_0=1.73$, $\alpha=1.9017$ and $\beta=0.539 (eV)^{-1}$ are the constant parameters for a given temperature and pressure. The temperature and pressure dependence of these parameters is again a subject of intensive future investigations.  Even though the above relation can be applied to semiconductors with a wide range of energy gaps, the prominent region in which this relation works well is within the range $0 < E_g < 5 eV$. Besides the Tripathy relation, some other popular empirical relations available in literature are 

\vspace{0.3cm}
Moss relation \cite{Moss50}:
\begin{equation}
n^4E_g=95 eV,\label{e2}
\end{equation}

\vspace{0.3cm}
Ravindra relation\cite{Ravi79}:
\begin{equation}
n=4.084-0.62E_g,\label{e3}
\end{equation}
\vspace{0.3cm}

Herve-Vandamme (HV) relation\cite{Herve94, Herve95}:
\begin{equation}
n^2=1+\left(\frac{A}{E_g+B}\right)^2,\label{e4}
\end{equation}
where $A$ is the hydrogen ionization energy $13.6eV$ and $B=3.47eV$ is a constant assumed to be the difference between UV resonance energy and band gap energy. Since II-VI and III-V groups of semiconductors have enough potential for optoelectronic device applications, we are interested in the present work to use the values of refractive indices calculated from these formulations to investigate some of their optical and electronic properties. For the purpose of calculation, we have considered those semiconductors belonging to these specific groups with energy gap lying in the range $0 < E_g < 5 eV$. The calculated values of refractive indices of these semiconductors can be obtained from Ref.\cite{SKT15}. 

\subsection{\it Dielectric constant and Linear optical susceptibility}

The dielectric properties of a material is usually measured through a frequency dependent complex dielectric function having its real and imaginary parts, $\epsilon (\omega)= \epsilon_1(\omega)+ i \epsilon_2(\omega)$. The dielectric function describes the response of the semiconductor to the electromagnetic radiation mediated through the interaction of photons and the electrons. It depends upon the electronic band structure and explains the combined excitations of Fermi sea. The real part of the dielectric constant determines the refractive index of a material and the imaginary part determines the absorption coefficient. The static limit dielectric constant $\epsilon_{\infty}$ is related to the refractive index $n$ through a simplified expression $\epsilon_{\infty}= n^2$, where the magnetic permeability of the medium is close to 1. In the present work, we have used the Tripathy relation in eq. \eqref{e1} to calculate the static limit of the dielectric constants of some II-VI and III-V semiconductors spanning over a wide range of energy gaps. In this context, we have considered semiconductors both from the narrow as well as wide band gaps. Once the dielectric constant is calculated, the linear optical susceptibility can then be obtained in a straightforward manner using  the relation $\chi = \epsilon_{\infty}-1$. In the II-VI group semiconductors, the dielectric constants are calculated for CdS, CdSe, ZnS, ZnTe, ZnSe, ZnO, MgS, MgTe, MgSe, SrS, SrTe, SrSe and are plotted as function of energy gaps in Figure 1. These semiconductors have energy gap lying in the range $1.5<E_g<5 eV$. In the figure, the known values of dielectric constants have also been shown for comparison. Our results are compared with the values calculated from the well known expressions of refractive index as proposed by Moss in eq.\eqref{e2}, Ravindra in eq.\eqref{e3} and Herve and Vandamme in eq.\eqref{e4}. It can be observed from the figure that, dielectric constant decreases with the increase in energy gap. The theoretical trends of the calculated dielectric constants fairly match with the experimental trend. However, in the lower energy gap region i.e. $E_g <2 eV$, the calculated values remain above the known ones and in the higher band gap region i.e $E_g>3.5eV$, the predicted values remain below the known dielectric constants. It is interesting to note from the figure that,  the theoretically calculated values from Tripathy relation agree fairly well with the known values at more number of data points as compared to that of HV relation particularly for ZnTe, ZnSe, ZnO, ZnS,  MgTe and SrS.  For BaSe and MgSe the calculated dielectric constants from Tripathy relation exactly match with the known values. Tripathy relation and HV relation behave alike for higher energy gap region whereas in the lower energy gap region, Tripathy relation predicts higher value of dielectric constant as compared to HV relation. The over prediction in the low energy gap region can be attributed to the exponential decay nature of the refractive index with energy gap in Tripathy relation. However, refractive index from Tripathy relation becomes softer in the wide band gap region and compares well with that of the HV relation. It is worth to mention here that, for the specific group of semiconductors considered here, HV relation predicts reasonably good values of refractive indices compared to those of Moss' and Ravindra relation \cite{SKT15}. Calculations from Ravindra relation do not have good agreement with the known values and therefore, for II-VI group of semiconductors in the energy gap range $1.5 < E_g < 5 eV$ one must be careful while using this relation. It can be emphasized that, Moss relation and Tripathy relation predict the dielectric constants reasonably well over the range of band gap considered in the present work and they can be used to a satisfactory extent for this group of semiconductors.

\begin{figure}[h!]
\begin{center}
\includegraphics[width=1\textwidth]{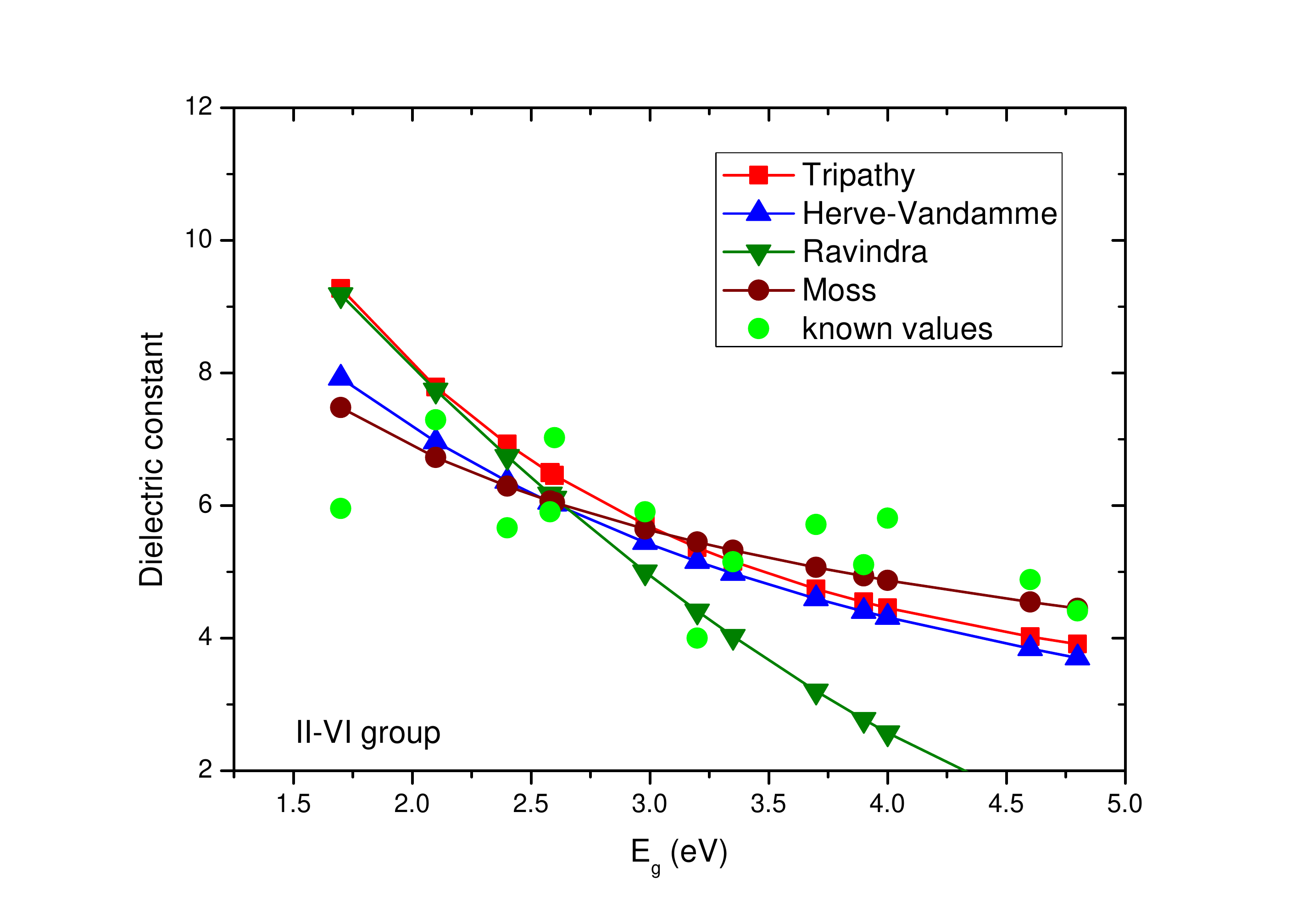}
\caption{Dielectric constant of II-VI group semiconductors as function of energy gap. The experimental values are shown as unconnected solid green dots.}
\end{center}
\end{figure}

In Figure 2, the dielectric constants of some III-V group semiconductors such as nitrides, phosphides, antimonides and arsenides of Ga, In and Al are shown along with the known values. The theoretical values are calculated using some popular relations including the Tripathy relation. The known values of dielectric constants are calculated using the known values of refractive indices from the relation $\epsilon_{\infty}=n^2$. For this specific group of semiconductors, dielectric constant clearly  shows a decreasing trend with energy gap. In fact, the decrement of the known values is linear with the energy gap. The theoretical calculations from Tripathy relation and others follow the experimental trend. Baring at few data points in the narrow band gap region i.e. $0<E_g<0.5 eV$, our calculations  have an excellent agreement with the known values. But it is interesting to note that, the dielectric constants calculated from Tripathy relation do not match with the known values for Indium based semiconductors such as InSb, InP and InAs. Our calculation over predicts the value of static dielectric constant for InSb, InP and InAs.  For Nitrides including that of InN, dielectric constants from Tripathy relation are in close agreement with the known values and the agreement is much better than the calculations from other relations. Also for Gallium based semiconductors, the agreement with the known values are excellent.  It is quite natural to expect that, in the low energy gap region, the calculated values from Tripathy relation become more stiff compared to that of Herve-Vandamme. Similar trend is also shown by Moss relation at narrow gap region. However, at high energy gap region i.e. $2<E_g<4 eV$, the theoretical calculations from all empirical relations except the Ravindra relation behave alike and also they agree with known values. In the high energy gap region, dielectric constants from Ravindra relation remain below the known values and those of other calculations. Moreover, for the prediction of static dielectric constant for III-V group semiconductors, Moss relation is a poor fit and can be reliable only in the region $ E_g>2.24 eV$, more specifically for nitrides of In, Ga and Al.

Since, the linear optical susceptibility is calculated from the relation $\chi = \epsilon_{\infty}-1$, it shows a similar trend as that of the dielectric constant plotted in figures 1 and 2 and therefore we can draw similar conclusions on the behaviour of $\chi$ for these group of semiconductors. For II-VI  group semiconductors, in the low energy gap region, calculations from Moss relation and HV relation are in closer agreement with that of the known ones and in the high energy gap region, Tripathy relation and Moss relation predict substantially accurate values of $\chi$. Except for the Zinc based compounds such as ZnTe, ZnSe and ZnO, Ravindra relation fails to predict the values of $\chi$ to satisfactory extent which may be due to the rapid linear decrement of refractive index with energy gap. For III-V group semiconductures, the predicted values of $\chi$ from Tripathy relation for all other semiconductors in this group are in excellent agreement with the known values except for InSb and InAs.

\begin{figure}[h!]
\begin{center}
\includegraphics[width=1\textwidth]{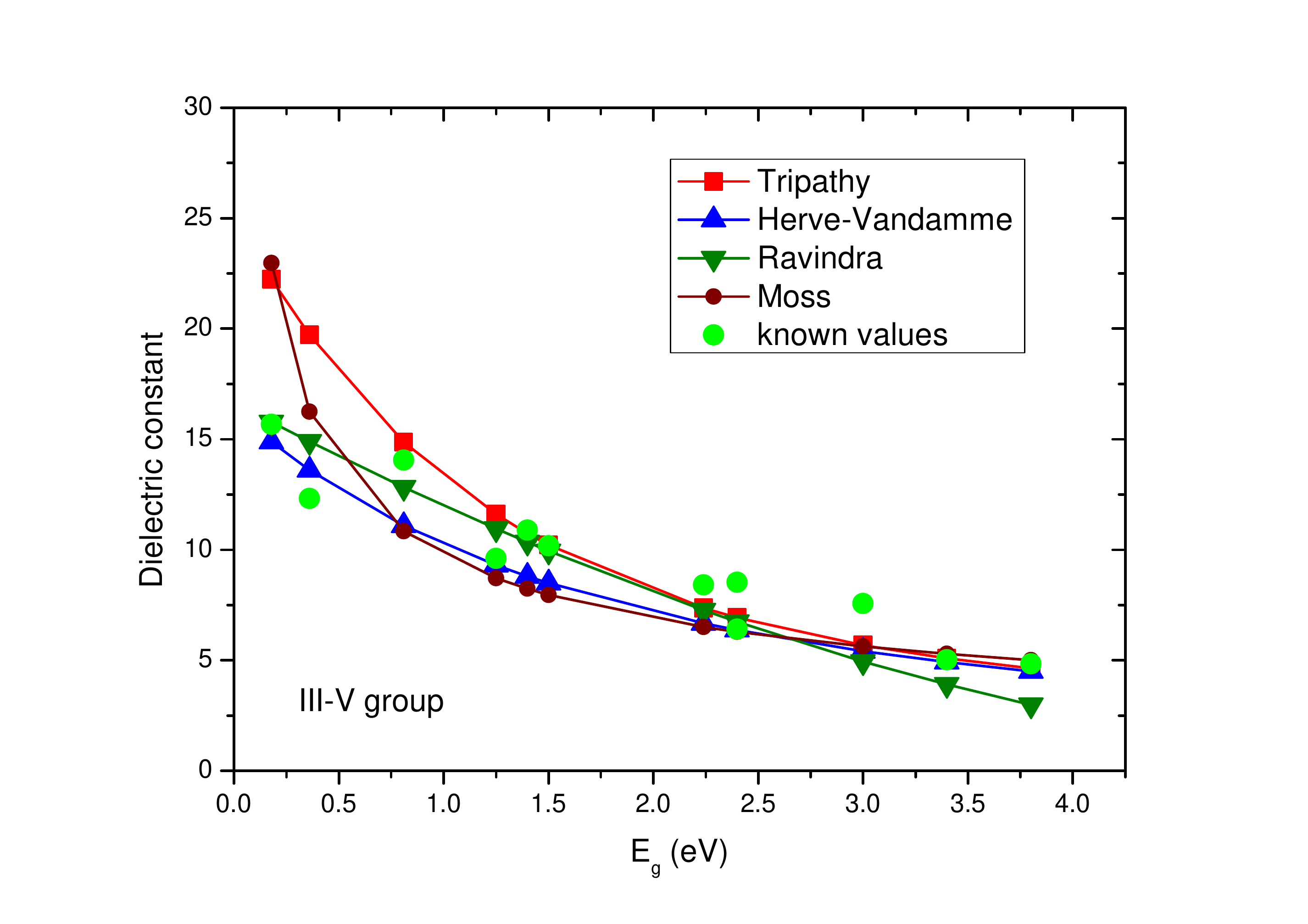}
\caption{Dielectric constants of III-V group semiconductors as function of energy gap. The experimental values are shown as unconnected solid green dots.}
\end{center}
\end{figure}

\subsection{\it Reflectivity}
Reflectivity $R$, defined through the ratio of the reflected power to incident power describes the optical response of the surface of a material. The reflectivity or reflection coefficient is calculated from the refractive index $n$ of a material using the expression 
\begin{equation}
R=\frac{\left(n-1\right)^2+k^2}{\left(n+1\right)^2+k^2},\label{e5}
\end{equation}
where, the extinction coefficient $k (\omega)$ is frequency dependent. We may assume a very small $k(\omega)$ for weakly absorbing media. Also, $k (\omega)$ vanishes at very high frequency i.e $k\rightarrow 0$. In such a situation, the reflectivity of a material can be calculated from the high frequency refractive index using the relation $R=\left(\frac{n-1}{n+1}\right)^2$.

\begin{figure}[h!]
\begin{center}
\includegraphics[width=1\textwidth]{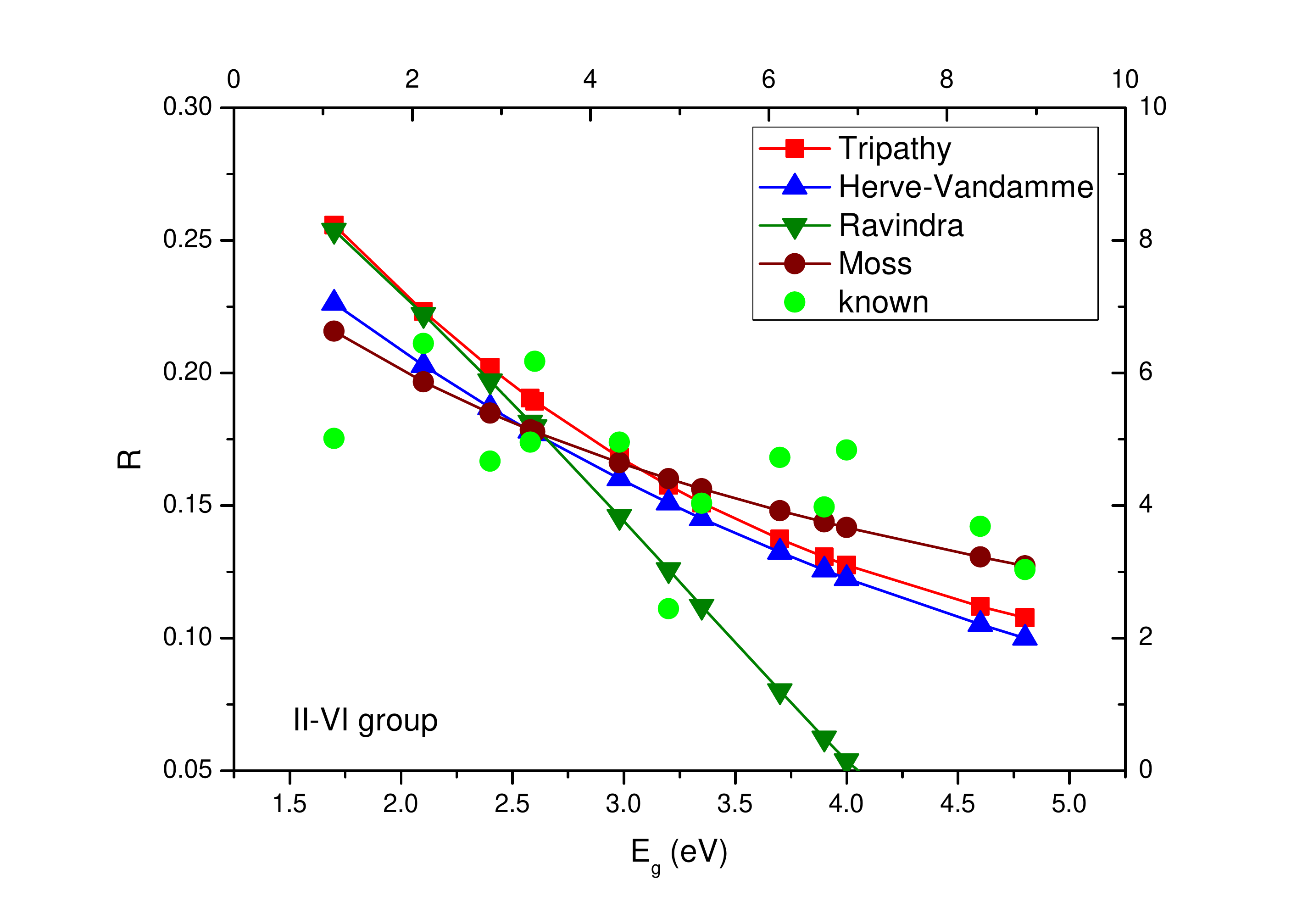}
\caption{Reflectivity of II-VI group semiconductors as function of energy gap. The experimental values are shown as unconnected solid green dots.}
\end{center}
\end{figure}

In the Figure 3, we have plotted the theoretically calculated values of reflectivity of some II-VI group binary semiconductors using the empirical relations for refractive index.  In general, reflectivity of materials decrease with the energy gap and fairly follows a linear trend. However, for ZnO, the reflectivity lies much below the theoretical values as predicted by the empirical relations including the Tripathy relation. It is interesting to note that, Ravindra relation predicts somewhat a closer value for the reflectivity of ZnO. For the of calculation of reflectivity of II-VI materials, Ravindra relation can not be reliable except for the ZnO, ZnTe and MgSe. Moss relation is more reliable compared to other cases for the calculation of reflectivity for II-VI group materials.

In Figure 4, the predicted reflectivity for some III-V group semiconductors are shown. In the figure, we have also shown the reflectivity of these semiconductors as calculated from the known values of refractive index. For this group of semiconductor, the known reflectivity follows a linear relationship with the energy gap with a negative slope parameter. Except in the low energy gap region in the range of  $E_g <0.5 eV$, the theoretical calculations from Tripathy relation have an excellent agreement with the known ones as compared to other calculations. The reason behind the non agreement with the known values in the low energy gap lies in the fact that, Tripathy relation has an exponential decrement of refractive index with respect to band gap and at low band gap region, the refractive index shows a higher slope compared to others. However, a similar trend is also observed for Moss relation at low energy gap region, where the refractive index behaves like $(E_g)^{-0.25}$. Ravindra relation can be reliable in the range $E_g < 2.2 eV$. In high energy gap region, calculations from Herve-Vandamme follow a similar trend with that of the Tripathy relation. This similar behaviour can be observed for all other optical parameters considered in the text. This is because, the refractive index of Tripathy relation behaves almost that of HV in the high gap region.

\begin{figure}[h!]
\begin{center}
\includegraphics[width=1\textwidth]{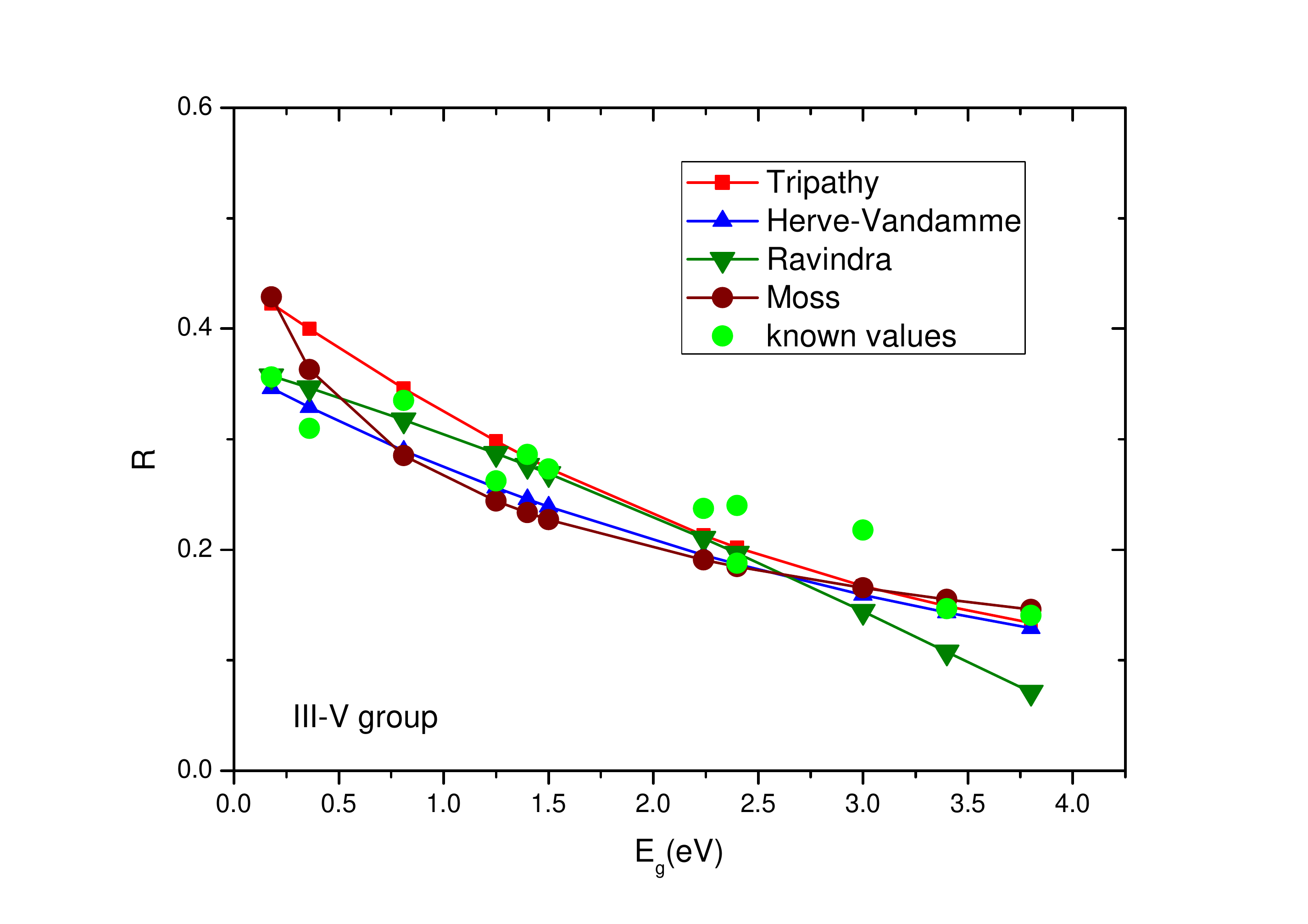}
\caption{Reflectivity of III-V group semiconductors as function of energy gap. The experimental values are shown as unconnected solid green dots.}
\end{center}
\end{figure}

\section{Electronic properties}

\subsection{\it Ionicity}
The group of binary II-VI and III-V semiconductors are materials with total of eight valence electrons, tetrahedrally bonded compound semiconductors. The electronic structures, charge distributions and phase stability as determined by the competition between the covalent $sp^3$ bonding and the electrostatic interaction play important roles in the observable properties of these compound semiconductors. In general, in ionic crystals, bonding between two atoms occur through the transfer of electron from one atom to the other resulting in two closed shell ions which interact through Coulomb force and a short range repulsive force described by certain potential. For covalent crystals there is no such charge transfer rather a kind of charge sharing occurs between the atoms. In the later case, the interactions are described through different chemical or physical processes of hybridisation among the orbitals. In this context, the origin of bonding configuration is explained through a parameter $f_i$ called ionicity. The ionicity of a bond is defined as the fraction $f_i$ of ionic or heteropolar character compared to the covalent or homopolar fraction $f_h$ so that $f_i+f_h =1$ \cite{Adachi05}. The range of this parameter is defined in the scale  $0\leq f_i \leq 1$ where $f_i =1$ is the ionic extreme \cite{Cart83}. After the thermochemical approach of Pauling\cite{Pauling32, Pauling60}, there have been many attempts to understand and to calculate this parameter relevant for the studies of bond and structure calculations in solid state Physics. From a molecular spectroscopic picture of bonding and antibonding states separated by an energy gap $E_g$, Phillips defined the ionicity scale $f_i$ for the tetrahedrally coordinated binary compounds $A^NB^{VIII-N}$ in terms of the homopolar $E_h$ and the heteropolar $C$ components of the complex band gap $E_g=E_h+iC$ as \cite{Adachi05, Phillips69, Phillips70, Phillips73, Garcia93}

\begin{equation}
f_i=\frac{C^2}{E_h^2+C^2}.\label{e6}
\end{equation}
The homopolar gap $E_h$ of the crystal is interpreted to be produced by the symmetric part of the total potential and the heteropolar gap $C$ results from the antisymmetric part of the potential. The homopolar gap can be calculated from a scaling of the optical gaps of group IV materials like diamond or silicon from the relation \cite{Garcia93}

\begin{equation}
E_h = E_h(Si)\left(\frac{a(Si)}{a}\right)^{2.5},\label{e7}
\end{equation}
where $a$ is the lattice constant. Phillips and Pauling scales of ionicity are widely used for the study of band structure and bonding configuration of compound semiconductors. One of the interesting feature of the dielectric theory of Phillips and Vechten \cite{Phillips69, Phillips70} is that, it is possible to define a critical ionicity value,  $f_i^c = 0.786$ to each compound that separates four fold-coordinated crystal structures (coordination number $N_C=4$) from sixfold-coordinated crystal structures (coordination number $N_C = 6$). After the proposal of Phillips and Pauling ionicity scales, there have been a lot  of interest in the bond ionicity of compounds \cite{Tubb70, Sharma83, Kumar87, Kumar87a, Christ87, Garcia93, AlDouri01, AlDouri10, AlDouri11, Kumar13}.

Assuming a nearly isotropic version of free electron model, Penn proposed a wave number dependent model for the calculation of dielectric function for semiconductors \cite{Penn62} which depends only on the average energy gap $E_g$ of the semiconductor. According to the Penn model, the high frequency dielectric constant is given by 

\begin{equation}
\epsilon_\infty =1+\left(\frac{\hbar \omega_p}{E_g}\right)^2 \left[1-\bigtriangleup +\frac{1}{3}\bigtriangleup^2\right],\label{e8}
\end{equation}
where $\bigtriangleup =\frac{E_g}{4E_F}$, $\hbar \omega_p$ is the valence electron plasmon energy and $E_F$ is the Fermi energy. 
The valence electron plasmon energy can be calculated from the refractive index of the material using an empirical relationship $\hbar \omega_p = k_1e^{-k_2 n}$, where the constants pair $(k_1, k_2)$ are $(22.079, 0.1779)$ and $(47.924,0.3546)$ respectively for II-VI and III-V group semiconductors \cite{Reddy03}. This relation has been successfully applied to a good number of binary semiconductors to predict their plasmon energies even though a direct method of calculation from the effective number of valence electrons taking part in plasma oscillations is available in literature \cite{Jackson78, Ravindra81}. The Fermi energy is evaluated from the relation $E_F = 0.2947 \left(\hbar \omega_p\right)^{\frac{4}{3}}$ \cite{Ravindra81}.

\begin{figure}[h!]
\begin{center}
\includegraphics[width=1\textwidth]{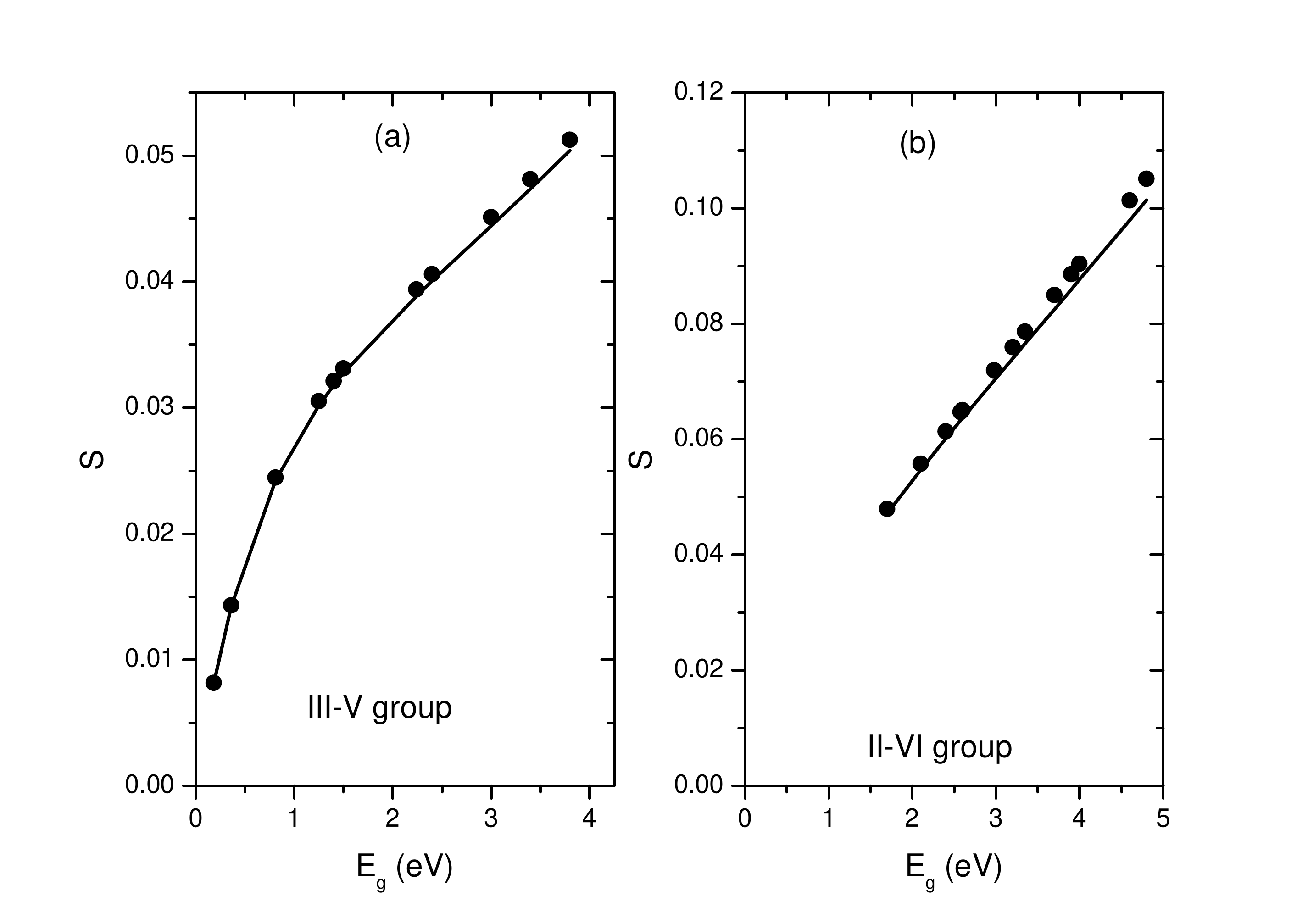}
\caption{ The deviation factor $S$ from unity is plotted as a function of energy gap. (a) For III-V group semiconductors (b) For II-VI group semicondutors. The calculated values of $S$ are shown by the solid line, whereas the solid dots are the values for the approximated $S\backsimeq \bigtriangleup$.}
\end{center}
\end{figure}

Tubb defined the crystal ionicity in terms of average energy gap $E_{av}$ and valence plasmon energy $\hbar \omega_p$ as \cite{Tubb70}

\begin{equation}
f_i^T = \frac{E_{av}}{\hbar \omega_p}.\label{e9}
\end{equation}
Using the fact that $\epsilon_{\infty}= n^2$ and $E_{av}=E_P$ (the Penn gap), from eq. \eqref{e8} and eq. \eqref{e9}, one can get a simplified relation for the ionicity in terms of the refractive index of materials as

\begin{equation}
f_i^T=\left[\frac{1-S}{n^2-1}\right]^{\frac{1}{2}}, \label{e10}
\end{equation}
where, $S = \left[\bigtriangleup -\frac{1}{3}\bigtriangleup^2\right] $ is a correction factor and marks the amount of deviation from unity. In his work, Penn, assumed this factor to be zero because the quantity $\bigtriangleup =\frac{E_g}{4E_F}$ can be negligibly small. In fact, for some elemental semiconductors such as Si and Ge, $S \simeq 0$. However, it has a significant contribution for many other compound semiconductors \cite{Ravindra81}. Since $\bigtriangleup$ is a very small quantity, higher order in $\bigtriangleup$ can be neglected and the deviation factor $S$ can be taken as $S\backsimeq \bigtriangleup$. In Figures 5(a) and 5(b), the deviation factor $S$ is shown as function of average energy gap for III-V and II-VI group of semiconductors. The deviation factor $S$ exponentially increases with energy gap for III-V group materials and linearly increases for II-VI group materials. Higher the energy gap of the material, higher is the deviation for $1-S$ from unity. The solid dots in the figures 5(a) and 5(b) represent the values of $\bigtriangleup$. At low energy gap region, $S$ is very small and both the values, $S$ and $\bigtriangleup$, exactly match, but with the increase in the energy gap, there remains a little discrepancy, the values of $\bigtriangleup$ remaining slightly above $S$. Practically, the difference will have a very little effect in the calculation of ionicity. It is now interesting to note that, if one gets the energy gap of a tetrahedrally coordinated binary semiconductor, one can get its bond ionicity  through the calculation of refractive index. In the present work, we have an interest to apply the recently proposed refractive index-energy gap relation as in eq.\eqref{e1} (Tripathy relation) to calculate the refractive index and then from the calculated values of refractive index we intend to calculate the plasmon energy for some tetrahedrally bonded binary III-V and II-VI group semiconductors. The plasmon energy will be used to get the Tubb ionicity. The calculated bond ionicities for III-V and II-VI group semiconductors are reported in  Table 1 and Table 2 respectively. In the tables, the Phillips scale ionicity are also given for comparison.
\begin{figure}[h!]
\begin{center}
\includegraphics[width=1\textwidth]{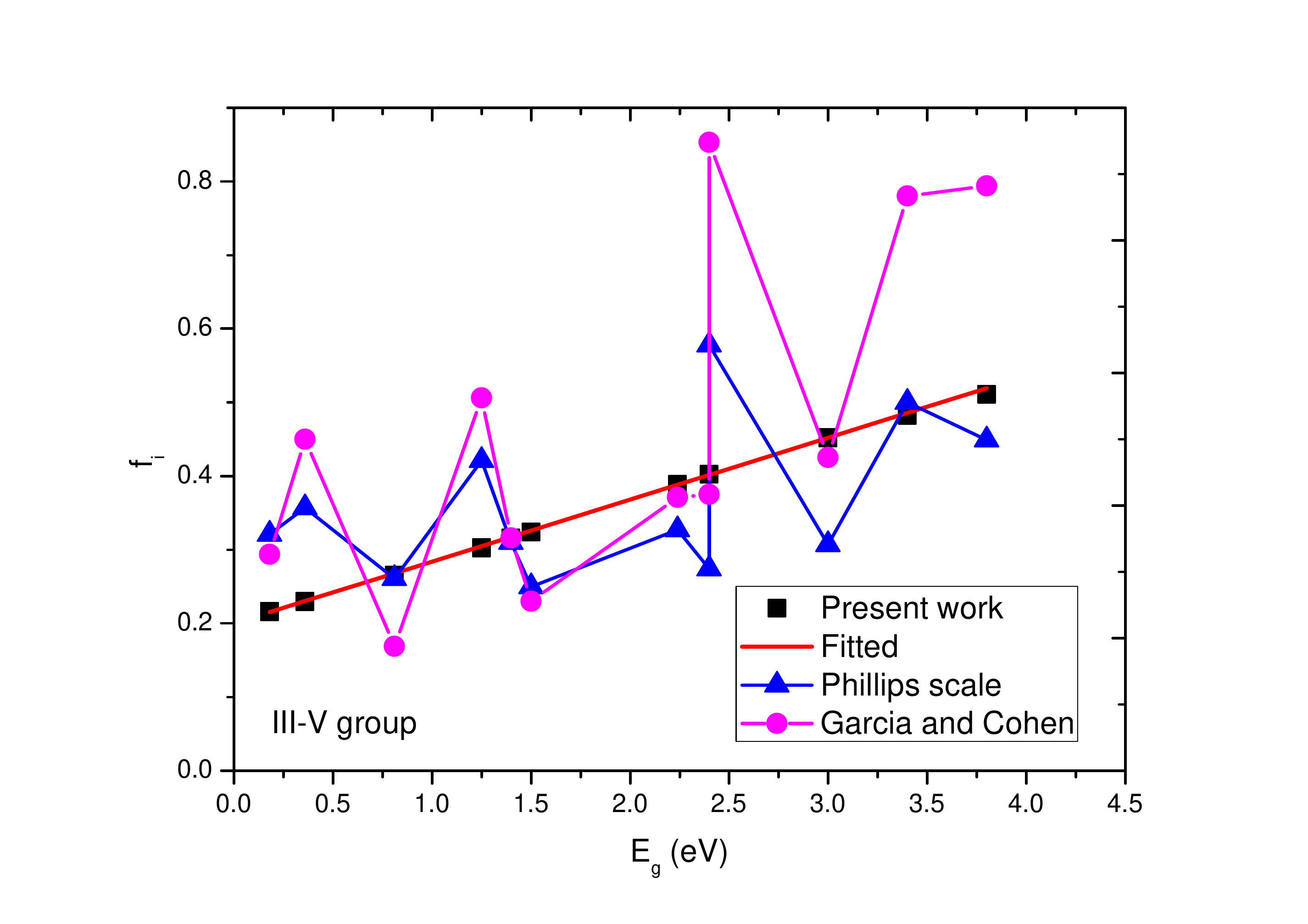}
\caption{Bond ionicity of III-V group semiconductors as function of energy gap.}
\end{center}
\end{figure}
\begin{figure}[h!]
\begin{center}
\includegraphics[width=1\textwidth]{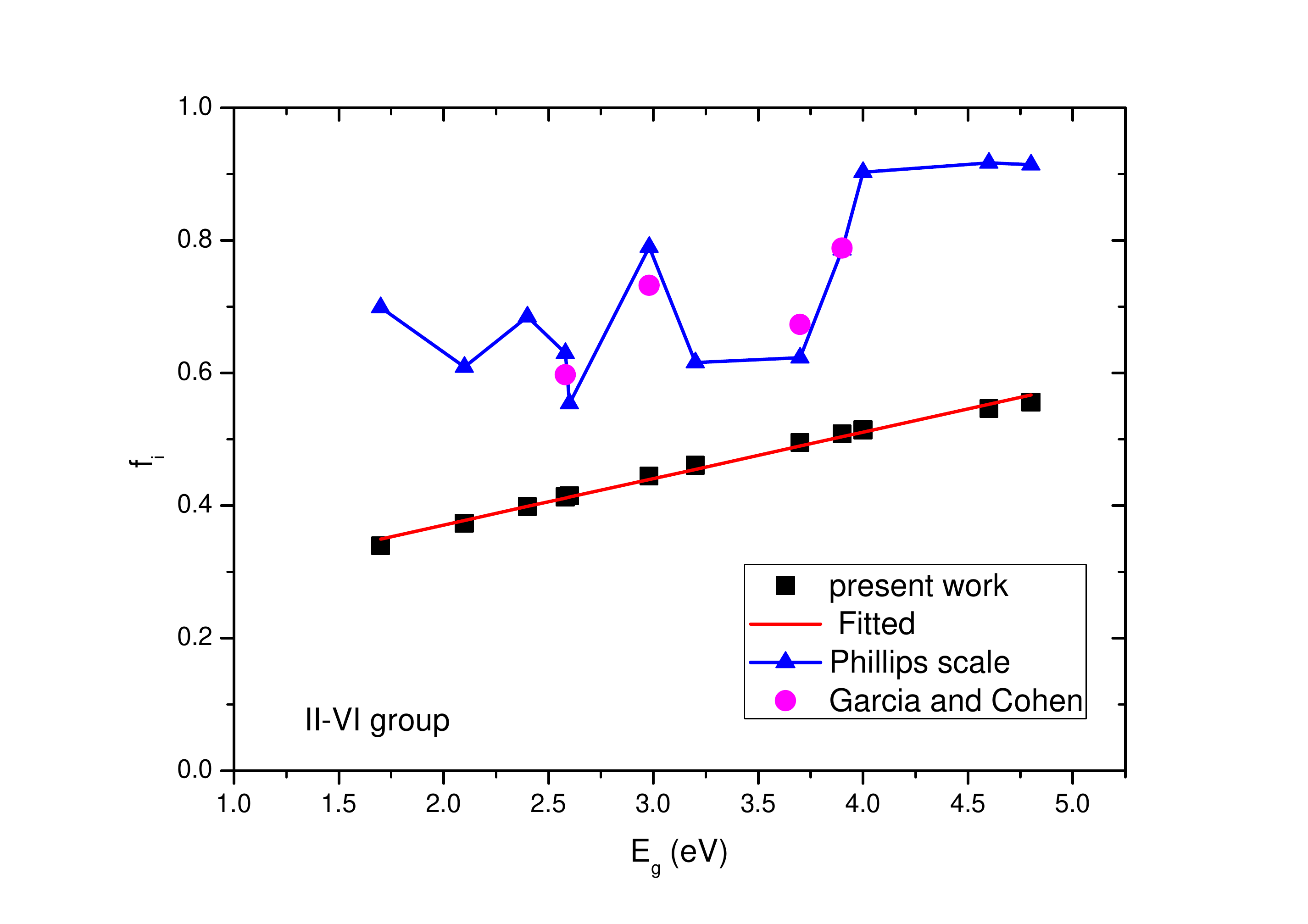}
\caption{Bond ionicity of II-VI group semiconductors as function of energy gap.}
\end{center}
\end{figure}


\begin{table}
\caption{Refractive index $n$, plasmon energy $\hbar \omega_p$, Fermi energy $E_F$ and ionicity $f_i$ of some III-V group semiconductors calculated from Tripathy relation.}
\centering
\begin{tabular}{l|c|c|c|c|c|c}
\hline \hline
Compounds	&	$E_g$ 	&n 	 		& $\hbar \omega_p$ & $E_F$		&$f_i^T$			&Phillips scale $f_i$\\
			&	($eV$)	&			&	$(eV)$		   & $(eV)$		&present work	& Ref.\cite{Phillips73, Garcia93}\\
\hline
$InSb$		&	0.18	&	4.72 	& 9.00				&	5.52	&0.216			&0.321\\
$InAs$		&	0.36	&	4.44	& 9.93				&	6.29	&0.23			&0.357\\
$GaSb$		&	0.81	&	3.86	& 12.21				&	8.29	&0.265			&0.261\\
$InP$		&	1.25	&	3.41	& 14.32				&	10.24	&0.302			&0.421\\
$GaAs$		&	1.4		&	3.28 	& 14.99				&	10.89	&0.315			&0.31\\
$AlSb$		&	1.5		&	3.19	& 15.43				&	11.32	&0.324			&0.25\\
$GaP$		&	2.24	&	2.71 	& 18.31				&	14.22	&0.389			&0.327\\
$AlAs$		&	2.4		&	2.63	& 18.84				&	14.78 	&0.402			&0.274\\
$InN$		&	2.4		&	2.63	& 18.84				&	14.78	&0.402			&0.578\\
$AlP$		&	3		&	2.38	& 20.59				&	16.63	&0.452			&0.307\\
$GaN$		&	3.4		&	2.26	& 21.53				&	17.65	&0.483			&0.5\\
$AlN$		&	3.8		&	2.15	& 22.32				&	18.53	&0.511			&0.449\\
\hline
\end{tabular}
\end{table}

\begin{table}
\caption{Refractive index $n$, plasmon energy $\hbar \omega_p$, Fermi energy $E_F$ and ionicity $f_i$ of some II-VI group semiconductors calculated from Tripathy relation.}
\centering
\begin{tabular}{l|c|c|c|c|c|c}
\hline \hline
Compounds	&	$E_g$ 	&	n 		& $\hbar \omega_p$ & $E_F$		&$f_i^T$			&Phillips scale $f_i$\\
			&	($eV$)	&			&	$(eV)$		   & $(eV)$		&present work	&	Ref.\cite{Phillips73, Garcia93}			\\
\hline
$CdSe$		&	1.7		&	3.05 	& 12.84				&	8.86	&0.339			&0.699\\
$ZnTe$		&	2.1		&	2.79 	& 13.44				&	9.42	&0.373			&0.609\\
$CdS$		&	2.4		&	2.63	& 13.82				&	9.78	&0.398			&0.685\\
$ZnSe$		&	2.58	&	2.55 	& 14.03				&	9.97	&0.413			&0.63\\
$MgTe$		&	2.6		&	2.54 	& 14.05				&	9.99	&0.414			&0.554\\
$MgSe$		&	2.98	&	2.39	& 14.43				&	10.35	&0.444			&0.79\\
$ZnO$		&	3.2		&	2.32 	& 14.62				&	10.59	&0.461			&0.616\\
$BaSe$		&	3.35	&	2.27	& 14.74				&	10.65 	&0.471			&-\\
$ZnS$		&	3.7		&	2.18	& 14.99				&	10.89	&0.495			&0.623\\
$MgS$		&	3.9		&	2.13	& 15.11				&	11.01	&0.508			&0.786\\
$SrTe$		&	4		&	2.11	& 15.17				&	11.06	&0.514			&0.903\\
$SrSe$		&	4.6		&	2.01	& 15.45				&	11.34	&0.546			&0.917\\
$SrS$		&	4.8		&	1.98	& 15.53				&	11.42	&0.556			&0.914\\
\hline
\end{tabular}
\end{table}

The Tubb ionicity scale as calculated from eq. \eqref{e10} and using Tripathy relation for refractive index for some III-V group semiconductors are plotted as function of energy gap in Figure 6. In the figure, the Phillips ionicity scale and the asymmetric charge coefficient $g$ of Garcia and Cohen \cite{Garcia93} which is a measure of ionic character of chemical bond are also shown for comparison. One can note that, for some compounds, the Garcia and Cohen coefficient $g$ lies above the Phillips scale and for some compounds it remains below. However, the dependence of the ionicity scale with energy gap, in both the cases, follow a similar trend of a fairly linear increase with energy gap. It is worth to mention here that, the energy gap increases with the increase in the charge asymmetry or ionicity for a given lattice constant mostly due to a downward shift of the valence band. This particular feature can be understood within the framework of a modified Kronig-Penney model.  It is found in the present work that the Tubb ionicity as calculated increases exactly linearly with the increase in energy gap and the trend line passes almost through the middle of the data points of Phillips and Garcia-Cohen. It is interesting to note that, for Gallium semiconductors such as GaSb, GaAs, GaP and GaN, our  calculation agree excellently with the Phillips scale and a bit fairly for AlN and AlSb. Also, our calculated values for GaAs, GaP, AlAs and AlP agree well with the Garcia-Cohen coefficients. In view of the linear behaviour of the Tubb ionicity  with energy gaps, we have fitted the calculated values to a linear equation 

\begin{equation}
f_i^T (E_g)=f_0^T+f_1^T E_g, \label{e11}
\end{equation}
where, the constants appearing in the above equations are $f_0^T= 0.20016 \pm 0.00145$ and $f_1^T = 0.08307 \pm 0.00066 (eV)^{-1}$. The fitted curve is also shown in the figure (red line).

The ionicity values calculated for II-VI group semiconductors are shown in Figure 7. In the figure, available Phillips scale and Garcia-Cohen coefficients are shown for comparison. One can note the deviation of the calculated Tubb ionicity from that of the Phillips scale and Garcia-Cohen values. The calculated values lie much below the corresponding Phillips scales and Garcia-Cohen values. However, like the Phillips scale ionicity and Garcia-Cohen coefficient, the present calculation for ionicity of II-VI group semiconductors shows a linearly increasing trend with energy gap. The calculated values of ionicity can be fitted to a linear equation such as eq. \eqref{e11} with renewed values of the constants: $f_0^T= 0.23058 \pm 0.00652$ and $f_1^T=0.07003 \pm 0.00194 (eV)^{-1}$. In the figure 7, the red line is for the fitted curve as shown along with the calculated values. The discrepancy between the Tubb ionicity scale as calculated in the present work using Tripathy relation and that of the Phillips scale may be due to some of the facts that we have not carefully looked into. The reasons may be (i) in the present calculation, we have used some empirical relations which may not be full proof for all groups of semiconductors and in all domain of energy gaps , (ii) the Tubb ionicity as has been defined may not have the real quantum picture involved in toto so that it differs with large magnitude when compared to others. However, as pointed out in Ref. \cite{Cart83}, Phillips ionicity scales fulfil a useful function in parametrising experimental data related to charge distribution but are questionable to the extent to which these scales provide useful information concerning the development of interatomic potentials and their use in solid state Physics. Above all, ionicity is closely related to the energy gap and vice versa. From the calculations of the ionicity of II-VI group and III-V group semiconductors, it is certain that, more or less, ionicity scale increases linearly with the energy gap. More is the ionicity, more can be the energy gap of the material, may it be a II-VI group semiconductor or be a III-V one.

\subsection{\it Electronic polarisability}
The response of a dielectric material to an external electric field is usually measured through the charge separation induced due to the external field which ultimately induces some dipole moment. The electronic polarisability, $\alpha$, of an ion relates the induced dipole moment $\mu$ to the local electric field $E_L$ acting on the ion as $\mu = \alpha E_L$ and therefore is a measure of the susceptibility of the ionic charge distribution to the distortion effected by the local electric field. 

The Lorentz local field can be approximated to the average local field through the dielectric polarisation {$\textbf P$} as \cite{Hinch00}

\begin{equation}
\textbf{F}=\langle \textbf {E}_L \rangle+ L \langle\textbf{P}\rangle,
\end{equation}
where $L$ is the dimensionless Lorentz factor depending on the structure of the material phase and can be equated to $\frac{4\pi}{3}$. The  Lorentz-Lorenz formula for refractive index can then be expressed as

\begin{equation}
\frac{n^2-1}{n^2+2}=\frac{4\pi N \alpha}{3}.
\end{equation}
Here, $N$ is the number of dipoles per unit volume of the material. Since refractive index for semiconductor can be determined from energy gap, its electronic polarisability (in the unit of $\angstrom ^3$ ; $\angstrom = 10^{-8} cm$ ) can be determined using the Lorentz-Lorenz formula:

\begin{equation}
\alpha = 0.3965\left(\frac{n^2-1}{n^2+2}\right)\frac{M}{\rho}.\label{e14}
\end{equation}
$M$ and $\rho$ are respectively the molecular weight and mass density of the material. We have calculated the electronic polarisability of some binary III-V and II-IV group semiconductors employing the refractive indices calculated from the Tripathy relation and then compared our results with those calculated from Herve-Vandamme, Ravindra and Moss relations. Our results are shown in figures 8 and  9 for III-V and II-IV group semiconductors respectively. The known values of electronic polarisability are also shown in the figures for comparison.  The known values of electronic polarisability of different semiconductors are calculated from eqn.(14) using the known values of refractive indices.

\begin{figure}[h!]
\begin{center}
\includegraphics[width=1\textwidth]{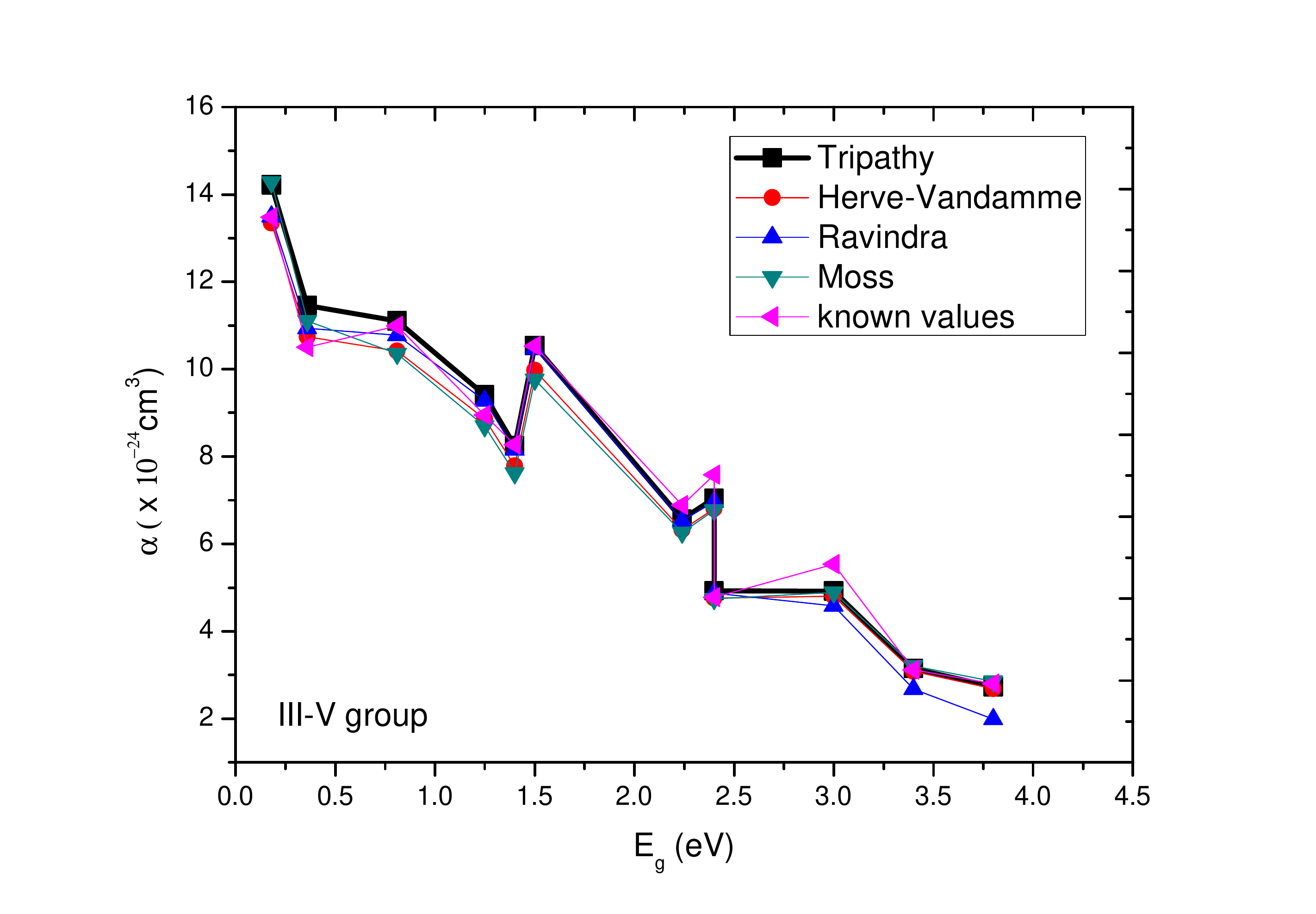}
\caption{The electronic polarisability of III-V group semiconductors as function of energy gap. The known values are calculated using the known refractive indices data collected from different reliable sources \cite{Kumar10, Weber03, Kasap07, Lide99}.}
\end{center}
\end{figure}
\begin{figure}[h!]
\begin{center}
\includegraphics[width=1\textwidth]{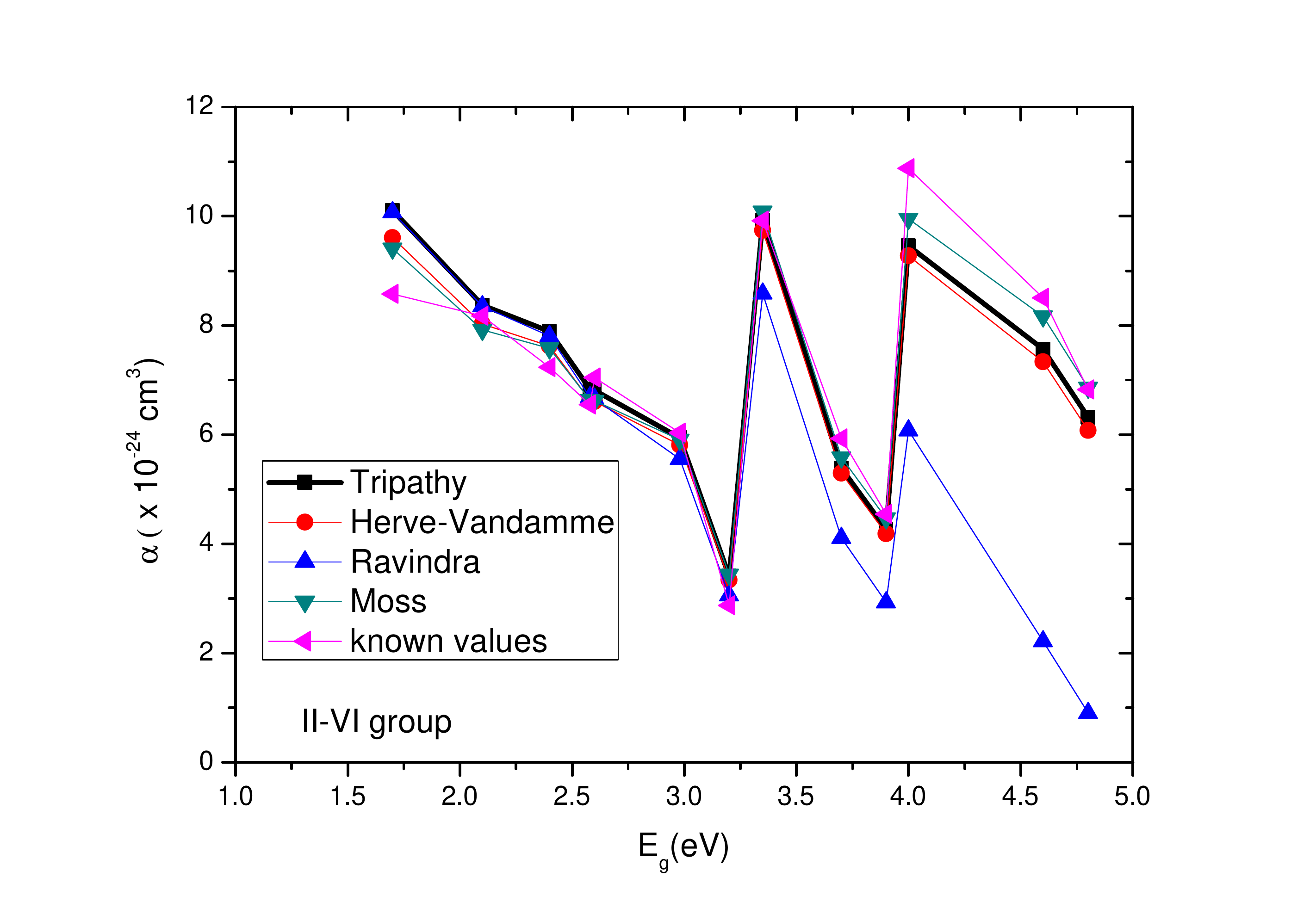}
\caption{The electronic polarisability of II-VI group semiconductors as function of energy gap. The known values are calculated using the known refractive indices data collected from different reliable sources \cite{Kumar10, Weber03, Kasap07, Lide99}.}
\end{center}
\end{figure}

The electronic polarisabilities for III-V group semiconductors, in general, decrease with energy gaps. However, this decrement is not very smooth. There occurs some bumps  for Aluminium based semiconductors such as AlSb, AlAs and AlP.  The calculated values of electronic polarisability from different relations almost follow the trend of the known values. Except the predictions from Ravindra relation, the calculated values differ a little from the known values. It is worth to mention here that, the Ravindra relation is not a good one for high energy gap region. Therefore it is obvious that, in the high energy gap region, the electronic polarisabilities calculated using this relation deviate much from the known values. One can note that, the calculated electronic polarisabilities using Tripathy relation reasonably match with the known values as compared to those calculated from other known relations. 

\begin{figure}[h!]
\begin{center}
\includegraphics[width=1\textwidth]{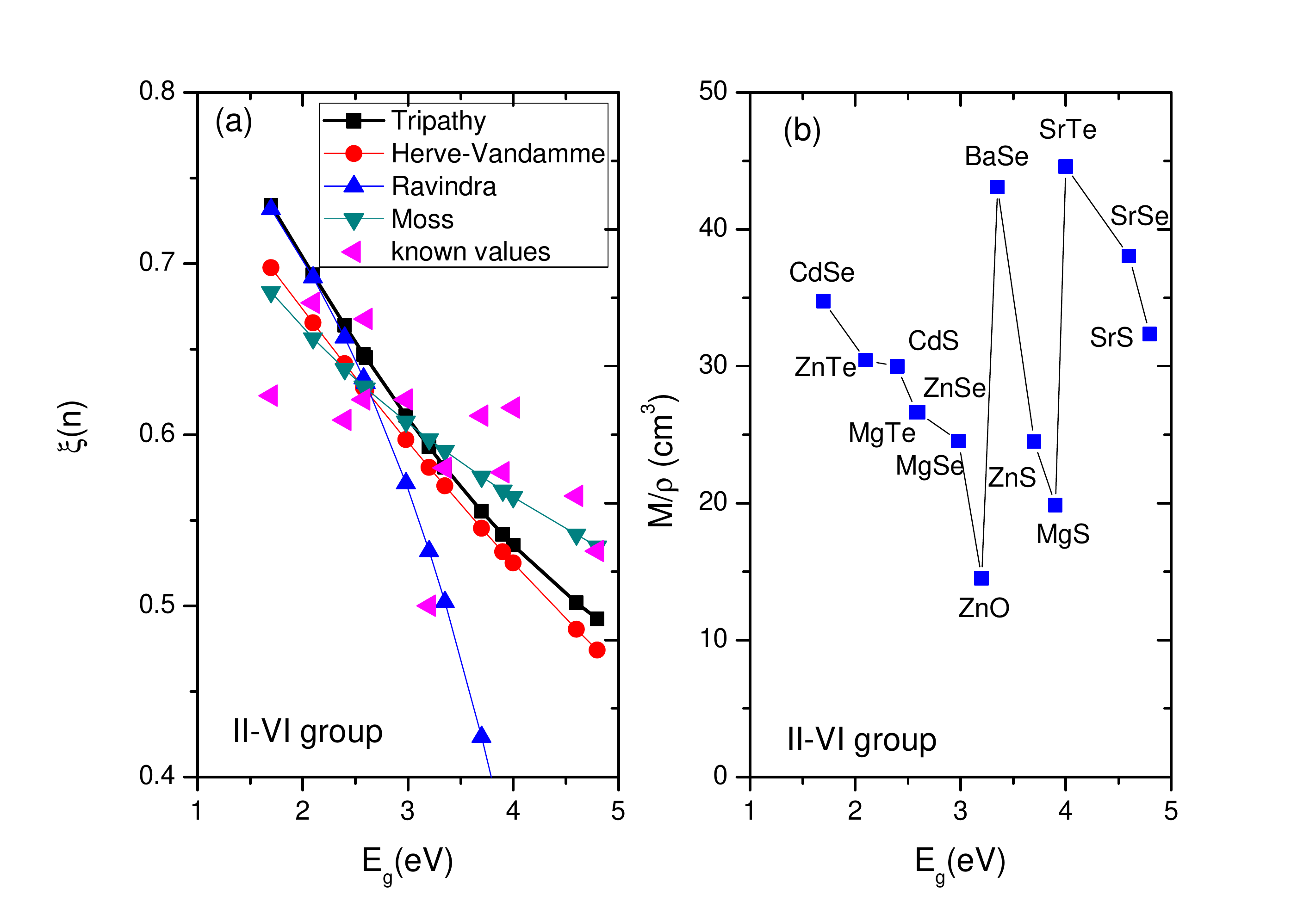}
\caption{(a) The functional $\xi (n) = \frac{n^2-1}{n^2+2}$ is plotted as a function of energy gap for II-VI group semiconductors. The known values are calculated using the known refractive indices data collected from different reliable sources \cite{Kumar10, Weber03, Kasap07, Lide99}. (b) The ratio $\frac{M}{\rho}$ for different II-VI group semiconductors arranged in an increasing order of energy gap.}
\end{center}
\end{figure}

The behaviour of electronic polarisability with energy gap for II-VI group semiconductors is obscure even though it decreases with energy gaps in the region $1.7 \leq E_g\leq 3.2 eV$ for CdTe, ZnTe, CdS, ZnSe, MgTe, MgSe and ZnO. For these binary semiconductors, the calculated electronic polarisabilities using different relations including Tripathy relation agree well with those calculated using known values of refractive indices. Similarly at high energy gap region $4\leq E_g \leq 4.8 eV$ for Sr based compounds, the electronic polarisability decreases with energy gap. However, if we consider the whole region of energy gaps from $1.7 eV $ to $4.8 eV$, there occur bumps for BaSe at $3.35 eV$ and for SrTe at $4 eV$. The electronic polarisability as in eq.\eqref{e14} has two major factors: the $E_g$ dependent part through the functional $\xi (n) = \frac{n^2-1}{n^2+2} $ and the ratio of molecular weight to mass density of the material $\frac{M}{\rho}$ besides the constant multiplicative factor. The roles played by both the factors have to be analysed in order to get an idea about the behaviour of electronic polarisability with respect to energy gap for II-VI group semiconductors. We have calculated these two factors for II-VI group semiconductors and shown them as function of energy gaps separately in figures 10(a) and 10(b). In figure 10(b), the ratio $\frac{M}{\rho}$ is shown for the selected semiconductors with increasing energy gap. It is clear from the figures that, the ratio $\frac{M}{\rho}$ has a dominant role in deciding the behaviour of the electronic polarisability than the functional $\xi (n)$. In fact, figure 9 appears to be a shadow of the figure 10(b). The functional $\xi (n)$ is $E_g$ dependent and smoothly decreases with the increase in energy gap. In view of the above discussion, it can be emphasized here that, the investigation of the behaviour of the electronic polarisability with energy gap, at least for II-VI semiconductors, should be carried out for specific compound semiconductors with a fixed element such as SrX, CdX, ZnX or MgX etc. In other way, instead of thinking of the whole expression as in eq.\eqref{e14}, one can consider the $E_g$ dependence of the functional $\xi (n)$ and then one should calculate the electronic polarisability. In order to get a clear $E_g$ dependence of the electronic polarisability of the binary semiconductors, we have fitted only the $E_g$ dependent part of eq.\eqref{e14} from the values calculated with known values of refractive indices of III-V and II-VI group semiconductors taken together to the form $\xi (E_g)=\frac{1}{a+ b E_g^{c}}$, where the constants $a, b$ and $c$ are determined from a best fit to the data. In the fitting procedure, these constants come out to be $a=1.1776$, $b=0.137 (eV)^{-1.0388}$ and $c=1.0388$. The electronic polarisability (in the unit of $\angstrom ^3$) of III-V and II-VI group binary semiconductors can now be expressed in terms of energy gap as

\begin{equation}
\alpha(E_g)= \frac{0.3965M}{\left(a+ b E_g^{c}\right)\rho}.\label{e15}
\end{equation}
The calculated electronic polarisabilities for III-V and II-VI group semiconductors from the fitted relation \eqref{e15} are given in tables 3 and 4 respectively. In the tables, the values of $\alpha$ calculated using Tripathy relation and Herve-Vandamme(HV) relation are also given for comparison. The values of $\alpha$ calculated from Lorentz-Lorenz relation using known values of $n$ are termed as known $\alpha$ in the tables. The calculated values of $\alpha$, particularly from Tripathy relation and the fitted relation agree well with the known values and are within acceptable accuracy. Also, the calculated and fitted $\alpha$ values are in close agreement with the values obtained in a recent work \cite{Aly15}. In the present investigation, we have emphasised that, if the energy gap of the compound semiconductor is known, then its electronic polarisability can be easily determined using equations \eqref{e1} and \eqref{e14}. Alternately, one can use equation \eqref{e15} to get the values of $\alpha$. In the event of non availability of energy gap value of a semiconductor, one can use the Duffy formulation \cite{Duffy80, Duffy90}, $E_g=3.72 \Delta x$ , to get the energy gap. Here $\Delta x$ is the electronegativity of the material. In terms of electronegativity $\Delta x$ of the material, the electronic polarisability can be expressed as 

\begin{equation}
\alpha(\Delta x)= \frac{0.3965M}{\left[a+ b \left(3.72 \Delta x\right)^{c}\right]\rho}.\label{e16}
\end{equation}
It is worth to mention here that, Duffy relation provides a good estimate of the energy gap with acceptable accuracy for materials whose electronegativity are known. In view of this, we believe that, equation \eqref{e16} can well be applied to a wide range of binary and ternary semiconductors which can be a subject for our future investigation.

\begin{table}
\caption{Electronic polarisability  ($\alpha$) of III-V group semiconductors.}
\centering
\begin{tabular}{l|c|c|c|c|c|c}
\hline \hline
Compounds	&	$E_g$ 	&n 	 		&\multicolumn{4}{c}{ Electronic polarisability $(\alpha) (\angstrom ^3)$}   \\
	\cline{4-7}
			&	($eV$)	&			&	 Known	 & Tripathy	& HV	        & Fitted (eq.\eqref{e15})\\
\hline
$InSb$		&	0.18	&	3.96 	& 13.48				&	14.22	&13.34			&13.52\\
$InAs$		&	0.36	&	3.51	& 10.51				&	11.46	&10.74			&10.85\\
$GaSb$		&	0.81	&	3.75	& 10.98				&	11.11	&10.41			&10.49\\
$InP$		&	1.25	&	3.1 	& 8.95				&	9.41	&8.86			&8.94\\
$GaAs$		&	1.4		&	3.3 	& 8.27				&	8.24	&7.79			&7.86\\
$AlSb$		&	1.5		&	3.19	& 10.53				&	10.54	&9.98			&10.08\\
$GaP$		&	2.24	&	2.9 	& 6.88				&	6.57	&6.32			&6.47\\
$AlAs$		&	2.4		&	2.92	& 7.58				&	7.04 	&6.80			&6.99\\
$InN$		&	2.4		&	2.53	& 4.77				&	4.93	&4.76			&4.89\\
$AlP$		&	3		&	2.75	& 5.53				&	4.91	&4.80			&5.02\\
$GaN$		&	3.4		&	2.24	& 3.12				&	3.14	&3.08			&3.27\\
$AlN$		&	3.8		&	2.2	    & 2.79				&	2.73	&2.68			&2.89\\
\hline
\end{tabular}
\end{table}

\begin{table}
\caption{Electronic polarisability  ($\alpha$) of II-VI group semiconductors.}
\centering
\begin{tabular}{l|c|c|c|c|c|c}
\hline \hline
Compounds	&	$E_g$ 	&n 	 		&\multicolumn{4}{c}{ Electronic polarisability $(\alpha) (\angstrom ^3)$}   \\
	\cline{4-7}
			&	($eV$)	&			&	 Known	 & Tripathy	& HV	        & Fitted (eq.\eqref{e15})\\
\hline
$CdSe$		&	1.7		&	2.44 	& 8.58				&		10.10	&		9.61	&	9.73\\
$ZnTe$		&	2.1		&	2.7 	& 8.17				&		8.37	&		8.03	&	8.19\\
$CdS$		&	2.4		&	2.38	& 7.23				&		7.98	&		7.62	&	7.83\\
$ZnSe$		&	2.58	&	2.43 	& 6.55	 			&		6.83	&		6.63	&	6.84\\
$MgTe$		&	2.6		&	2.65	& 7.05				&		6.81	&		6.61	&	6.82\\
$MgSe$		&	2.98	&	2.43	& 6.03				&		5.94	&		5.81	&	6.06\\
$ZnO$		&	3.2		&	2   	& 2.87				&		3.40	&		3.34	&	3.51\\
$BaSe$		&	3.35	&	2.27	& 9.91				&		9.92 	&		9.74	&	10.30\\
$ZnS$		&	3.7		&	2.39	& 5.93				&		5.38	&		5.29	&	5.67\\
$MgS$		&	3.9		&	2.26	& 4.55				&		4.26	&		4.18	&	4.52\\
$SrTe$		&	4		&	2.41	& 10.08				&		9.45	&		9.27	&	10.06\\
$SrSe$		&	4.6		&	2.21	& 8.50				&		7.56	&		7.33	&	8.17\\
$SrS$		&	4.8		&	2.1 	& 6.82				&		6.31	&		6.08	&	6.83\\
\hline
\end{tabular}
\end{table}

\section{Summary and Conclusion}
In the present work, we have studied some of the optical and electronic properties of II-VI and III-V group binary semiconductors with respect to their behavioural determination from energy gaps. The motivation behind the choice of these specific group of semiconductors is very clear: these semiconductors have enough potential for new device applications enriched with opto-electronic behaviour. Here, we emphasized upon the calculation of different optical and electronic properties of the semiconductors from their energy gaps. We have calculated the refractive indices from their energy gaps using some well known refractive index-energy gap relations available in literature including the recently proposed Tripathy relation \cite{SKT15}.  Different optical and electronic properties of these group semiconductors are then calculated from their respective refractive indices. 

From the present investigation it is found that, dielectric constant, linear optical susceptibility and reflectivity of II-VI and III-V group semiconductors decrease with the energy gap.  Even though, the calculations of refractive index from Tripathy relation and Herve-Vandamme relation are quite closer \cite{SKT15},  Tripathy relation agree with the known values at more number of data points than HV relation in the optical properties of these semiconductors. For II-VI group semiconductors, Moss' and Tripathy relation predict with reasonable accuracy the values of dielectric constant and linear optical susceptibility. Ravindra relation is a poor fit for this particular group. However, for Zinc based compounds Ravindra relation can be reliable. For III-V group semiconductors, calculations from Tripathy relation excellently agree with the known values but it fails to predict accurately the dielectric constant and linear optical susceptibility for Indium based compounds.

In the electronic properties of these specific groups of semiconductors, we have calculated Tubb ionicity scales and electronic polarisability.  The calculated  ionicity are also compared with the Phillips ionicity scale and Garcia-Cohen coefficient. Obviously, our calculation with the Tripathy relation, being an empirical one, does not match with others for II-VI semiconductors. However, for III-V group semiconductors, our calculation passes almost through the middle of the data points of Phillips and Garcia-Cohen. More interestingly, for Gallium based semiconductors our calculations are in excellent agreement with the Phillips scale. The ionicity scale even though provide useful information of the inter atomic potentials but their use may be questionable for use in solid state Physics. However, ionicity is closely associated with the energy gap and from our calculation, we observed that the Tubb ionicity scale increases linearly with the energy gap for both the group of semiconductors. 

Electronic polarisability, in general, is observed to decrease with energy gaps. The decrement for II-VI group is not much straightforward as is evident in III-V group semiconductors. The uneven behaviour of electronic polarisability for II-VI group is analysed. It is certain from the analysis that,  the structure and composition of the material in the form of the ratio of molecular weight to mass density plays a dominant role in the calculation of electronic polarisability than the contribution coming from refractive index. In view of this, we have suggested an empirical relation for its calculation. The calculated values from the empirical relation are in excellent agreement with the known values. Even though, the empirical relation is a simple one, we believe that, it can be applied for a wide range of binary as well as ternary semiconductors.

\end{document}